\documentclass[onecolumn,aps,prd,superscriptaddress,preprintnumbers,nofootinbib,10pt]{revtex4-2}
\usepackage{amsfonts}
\usepackage[tbtags]{amsmath}
\usepackage{amssymb}
\usepackage{color}
\usepackage{courier}
\usepackage{dsfont}
\usepackage{euscript}
\usepackage{epsfig}
\usepackage{epstopdf}
\usepackage{float}
\usepackage{graphics}
\usepackage{subcaption}
\usepackage{graphicx}
\usepackage[utf8]{inputenc}
\usepackage{latexsym}
\usepackage{MnSymbol}
\usepackage{pifont}
\usepackage{physics}
\usepackage{slashed}
\usepackage{xcolor}
\usepackage[normalem]{ulem}
\usepackage{url}
\usepackage{verbatim}
\usepackage{widetable}
\usepackage{setspace}
\usepackage[plainpages=false,pdfpagelabels]{hyperref}

\allowdisplaybreaks

\hypersetup{
	colorlinks=true,
	citecolor=blue,
	citebordercolor=red,
	linktoc=all,
	linkcolor=blue,
	urlcolor=blue
}

\definecolor{BLUE}{rgb}{0,0,1}

\begin{document}

		\title{\bf \Large Modular wedge localization, Majorana fields and the Tsirelson limit \\ of the Bell-CHSH inequality}
	
		\vspace{1cm}

	\author{J. G.A. Caribé}\email{joaogcaribe@ufrrj.br} \affiliation{ UFRRJ, Universidade Federal Rural do Rio de Janeiro, Departamento de Física, Zona Rural, BR-465, Km 07, 23890-000, Seropédica, Rio de Janeiro, Brazil}\affiliation{UERJ -- Universidade do Estado do Rio de Janeiro,	Instituto de Física -- Departamento de Física Teórica -- Rua São Francisco Xavier 524, 20550-013, Maracanã, Rio de Janeiro, Brazil}

	\author{M. S.  Guimaraes}\email{msguimaraes@uerj.br} \affiliation{UERJ -- Universidade do Estado do Rio de Janeiro,	Instituto de Física -- Departamento de Física Teórica -- Rua São Francisco Xavier 524, 20550-013, Maracanã, Rio de Janeiro, Brazil}
	
	\author{I. Roditi} \email{roditi@cbpf.br} \affiliation{CBPF $-$ Centro Brasileiro de Pesquisas Físicas, Rua Dr. Xavier Sigaud 150, 22290-180, Rio de Janeiro, Brazil}\affiliation{Institute for Theoretical Physics, ETH Z\"urich, 8093 Z\"urich, Switzerland} 
	
	\author{S. P. Sorella} \email{silvio.sorella@fis.uerj.br} \affiliation{UERJ -- Universidade do Estado do Rio de Janeiro,	Instituto de Física -- Departamento de Física Teórica -- Rua São Francisco Xavier 524, 20550-013, Maracanã, Rio de Janeiro, Brazil}

	\begin{abstract}

The massive Majorana field in $1+1$ dimension is employed to investigate the violation of the Bell-CHSH inequality in relativistic Quantum Field Theory. We give an explicit rapidity-space realization of the Summers-Werner modular-localization construction and reduce the vacuum Bell-CHSH correlator to a single spectral weight $h^2(\omega)$ for the modular operator. The resulting analytic families approach the Tsirelson bound in the vacuum state as their spectral weight concentrates near $\omega\approx0$, corresponding to the eigenvalue $\lambda^2 \approx 1$ of the modular operator.

\end{abstract}
		
\maketitle

\section{Introduction}\label{sect}

The study of the Bell-CHSH inequality \cite{Bell:1964kc,Clauser:1969ny} in relativistic Quantum Field Theory extends the corresponding question in Quantum Mechanics\footnote{See \cite{Guimaraes:2024byw,Guimaraes:2024mmp}
 for an introduction to the subject.}, with applications ranging from quantum information to the algebraic structure of entanglement in field theory \cite{Witten:2018zxz}. \\\\For wedge regions, the starting point is the work of Summers and Werner \cite{Summers:1987fn,Summers:1987squ}. They gave a rigorous formulation of Bell-CHSH tests in Quantum Field Theory and showed that the vacuum of a free relativistic field theory can display the maximal violation of a Bell-CHSH inequality allowed by the Tsirelson bound \cite{Cirelson:1980ry}. Their proof uses Haag-Kastler locality \cite{Haag:1992hx}, the Reeh-Schlieder theorem \cite{Reeh:1961ujh}, the Bisognano-Wichmann theorem for wedge regions \cite{Bisognano:1975ih}, von Neumann algebras \cite{Sorce:2023fdx}, Tomita-Takesaki theory \cite{Takesaki:1970aki,Bratteli:1979tw,Summers:2003tf,Guido:2008jk}, and modular localization \cite{Brunetti:2002nt,Borchers:2000pv,Schroer:2014kya}. \\\\The purpose of the present paper is more focused and more explicit. We consider the free massive Majorana field in $1+1$ dimensions and mplement the wedge-localized Bell-CHSH construction directly in rapidity space. The result is a closed formula for the vacuum Bell-CHSH correlator in terms of one spectral weight $h^2(\omega)$ associated with the modular operator. This gives a concrete model in which the approach to the Tsirelson bound can be followed analytically. \\\\The value $2\sqrt{2}$ is obtained as the supremum of a sequence of admissible spectral weights concentrating near $\omega=0$, equivalently near the eigenvalue of the modular operator $\lambda^2(\omega)=e^{-2\pi\omega}=1$. For regular normalizable weights the Bell-CHSH value remains below $2\sqrt{2}$, although it can be made arbitrarily close to it. \\\\Fermi fields are especially convenient for this question. The canonical anti-commutation relations give dichotomic Hermitian field operators after normalization \cite{Summers:1987fn,Summers:1987squ,Dudal:2023mij,Dudal:2026eil}.\footnote{Since odd Fermi fields are graded-local rather than local in the ordinary commuting sense, we shall use the standard fermionic twist, equivalently a Klein transformation on one wedge, when formulating the Bell-CHSH operator. This point is made explicit in Section \eqref{sect4}.} A fully analogous construction for Bose fields is less direct; see \cite{Caribe:2026mam} for a recent discussion. \\\\The work is organized as follows. In Section \eqref{sect2} we recall the quantization of the massive Majorana field in $1+1$ dimensions. Section \eqref{sect3} treats the smearing procedure and the 1-particle Hilbert space. Section \eqref{sect4} formulates the Bell-CHSH operator using Tomita-Takesaki theory, modular wedge localization, and the fermionic twist. In Section \eqref{sect5} we derive the spectral formula and give analytic families approaching the Tsirelson bound. Section \eqref{sect6} contains our conclusion.

\section{The massive Majorana field in $1+1$ and its quantization}\label{sect2}

The action of the model is specified by the following expression

\begin{equation} 
S = \int d^2x \; \left( \frac{i}{2} {\bar \psi} \gamma^\mu \partial_\mu \psi - \frac{m}{2} {\bar \psi}\psi \right)\;, \label{Mact}
\end{equation}
with 
\begin{equation} 
\gamma^0=\begin{pmatrix}
0 & 1 \\
1 & 0
\end{pmatrix} \;, \qquad
\gamma^1=\begin{pmatrix}
0 & 1 \\
-1 & 0
\end{pmatrix}\;, \qquad g_{\mu\nu}=diag(1,-1) \;. \label{conv}
\end{equation}
The field $\psi$ is a two component spinor fulfilling the Majorana condition 
\begin{equation} 
\psi^{C} = \psi \;, \qquad \psi^{C} = C {\bar \psi}^T \;,  \label{mj1}
\end{equation}
where $\bar \psi = \psi^{\dagger}\gamma^{0}$ and $ C$ is the charge conjugation matrix 
\begin{equation} 
C= \gamma^1 \;, \qquad C \gamma^\mu C^{-1}= {\color{red}{(\gamma^{\mu})^T}} \;, \qquad C^2=-1 \;. \label{chconj}
\end{equation} 
Condition \eqref{mj1} gives 
\begin{equation} 
\psi(t,x) = \begin{pmatrix}
h(t,x)  \\
i \varphi(t,x)
\end{pmatrix} \;, \label{mj2}
\end{equation}
with $(h,\varphi)$ real fields. From the equations of motion one gets 
\begin{equation} 
(\partial_t + \partial_x)\varphi =- m h  \;, \qquad (\partial_t - \partial_x) h = m \varphi  \;, \label{mj3}
\end{equation}
from which one sees that the two components $(h,\varphi)$ get mixed by the mass term. \\\\In order to write down the plane wave expansion of $\psi$, it is helpful to employ the rapidity variable $\theta$ 
\begin{equation} 
\omega_p = \sqrt{p^2+m^2} = m \cosh(\theta) \;, \qquad p = m \sinh(\theta) \;, \qquad \theta \in \mathbb{R} \;, \label{rap}
\end{equation}
so that 
\begin{equation} 
\frac{dp}{\omega_p} = d\theta \;. \label{rap1}
\end{equation}
For the plane wave expansion one obtains 
\begin{equation} 
\psi(t,x) = \int_{-\infty}^{\infty} \frac{d \theta}{2 \pi}\left( b(\theta) u(\theta) e^{-i p^\mu(\theta)x_\mu} + b^\dagger(\theta) v(\theta) e^{i p^\mu(\theta)x_\mu}  \right) \;, \qquad p^\mu(\theta)x_\mu = m(\cosh(\theta)t - \sinh(\theta)x ) \;, \label{wexp}
\end{equation}
where $(b(\theta), b^\dagger(\theta))$ are the annihilation and creation operators and $(u(\theta),v(\theta))$ the positiveand negative frequency solutions of the equations of motion given, respectively, by 
\begin{equation} 
u(\theta)  = \frac{1}{\sqrt{2}}\begin{pmatrix}
e^{-\frac{\theta}{2}}  \\
e^{\frac{\theta}{2}}
\end{pmatrix} \;, \qquad u(\theta) \otimes {\bar u}(\theta) = \frac{1}{2} \begin{pmatrix}
1 & e^{-{\theta}} \\
e^{{\theta}} & 1
\end{pmatrix} \;,\label{uth}
\end{equation}
and 
\begin{equation} 
v(\theta)  = \frac{1}{\sqrt{2}}\begin{pmatrix}
e^{-\frac{\theta}{2}}  \\
- e^{\frac{\theta}{2}}
\end{pmatrix} \;, \qquad v(\theta) \otimes {\bar v}(\theta) = \frac{1}{2} \begin{pmatrix}
-1 & e^{-{\theta}} \\
e^{{\theta}} & -1
\end{pmatrix} \;.\label{vth}
\end{equation}
Due to the Majorana condition  \eqref{mj1}, it turns out that 
\begin{equation} 
v(\theta) = C ( {\bar u}(\theta))^T  \;. \label{cmj}
\end{equation}
Therefore, for the components $(h,\varphi)$, one writes 
\begin{eqnarray} 
h(t,x) & = & \int_{-\infty}^{\infty} \frac{d \theta}{2 \pi} \frac{e^{-\frac{\theta}{2}}}{\sqrt{2}}  \left( b(\theta)  e^{-i p^\mu(\theta)x_\mu} + b^\dagger(\theta)  e^{i p^\mu(\theta)x_\mu}  \right) \;, \nonumber \\
\varphi(t,x) & = & \frac{1}{i}\int_{-\infty}^{\infty} \frac{d \theta}{2 \pi} \frac{e^{\frac{\theta}{2}}}{\sqrt{2}}  \left( b(\theta)  e^{-i p^\mu(\theta)x_\mu} - b^\dagger(\theta)  e^{i p^\mu(\theta)x_\mu}  \right) \;. \label{comp} 
\end{eqnarray}
With the above conventions, for the canonical anti-commutation relations, we have 
\begin{equation} 
\{ b(\theta), b^\dagger(\theta') \} = (2\pi) m \;\delta(\theta-\theta') \;, \qquad \{ b(\theta), b(\theta') \} = \{ b^\dagger(\theta), b^\dagger(\theta') \} =0 \;, \label{ccrb}
\end{equation}
from which one easily checks that 
\begin{equation} 
\{ h(t,x), h(t,y) \} = \delta(x-y) \;, \qquad \{ \varphi(t,x), \varphi(t,y) \} = \delta(x-y) \;, \qquad \{ h(t,x), \varphi(t,y) \} = 0 \;. \label{ccrf}
\end{equation}

\section{Smearing and 1-particle Hilbert space }\label{sect3}

As one learns from \cite{Haag:1992hx}, fields are operator valued distributions. As such, they need to be smeared in order to obtain well defined operators acting on the Hilbert space. In the present case, being $\psi$ a spinor field, the smearing is performed by a two component spinor test function 
\begin{equation} 
f(t,x) = \begin{pmatrix}
i f_2(t,x)  \\
f_1(t,x)
\end{pmatrix} \;, \label{ftest}
\end{equation} 
with $(f_1,f_2)$ real smooth functions. Setting ${\bar f}= f^\dagger \gamma^0$, for the smeared Majorana field $\psi(f)$ one gets 
\begin{equation} 
\psi(f) = \int d^2x\; {\bar f}(t,x) \psi(t,x) = \int d^2x \left(f_1(t,x) h(t,x) + f_2(t,x) \varphi(t,x) \right) = \psi(f)^\dagger \;. \label{sm}
\end{equation}
Introducing the on-shell Fourier transformation 
\begin{equation} 
f_j(\theta) = \int d^2x e^{i p^\mu(\theta)x_\mu} f_j(t,x) \;, \qquad j=1,2 \;, \qquad {\color{red}p(\theta)^\mu p(\theta)_\mu=m^2} \;, \label{ons}
\end{equation}
the smeared expression \eqref{sm} can be rewritten as 
\begin{equation} 
\psi(f) = b_f + b^\dagger_f \;, \label{sm1}
\end{equation}
with 
\begin{equation} 
b_f = \frac{1}{\sqrt{2}} \int_{-\infty}^{\infty} \frac{d \theta}{2 \pi}\left( f_1^*(\theta) e^{-\frac{\theta}{2}} - i f_2^*(\theta) e^{\frac{\theta}{2}}\right) b_{\theta}\;, \qquad b_f^\dagger = \frac{1}{\sqrt{2}} \int_{-\infty}^{\infty} \frac{d \theta}{2 \pi}\left( f_1(\theta) e^{-\frac{\theta}{2}} + i f_2(\theta) e^{\frac{\theta}{2}}\right) b_{\theta}^\dagger\;, \label{bsm}
\end{equation}
denoting the smeared annihilation and creation operators. One feature which should be observed about expressions \eqref{bsm} is that the components $(f_1(\theta),f_2(\theta))$ of the spinor test function enter through the special combination 
\begin{equation} 
{\hat f}(\theta) = \left( e^{ -\frac{\theta}{2}} f_1(\theta) + i e^{ \frac{\theta}{2}} f_2(\theta) \right) \;. \label{comb}
\end{equation}
The component $f_1(\theta)$ is multiplied by the weight $e^{ -\frac{\theta}{2}}$, while $f_2(\theta)$ by $e^{ \frac{\theta}{2}}$. These are precisely the weights entering the $u(\theta)$ and $v(\theta)$ spinors in expressions \eqref{uth},\eqref{vth}. Reminding now that, under a boost transformation $\Lambda_s$ with parameter $s$ 
\begin{equation} 
\omega'_p = \omega_p \cosh(s) + p \sinh(s) \;, \qquad p'=p \cosh(s) + \omega_p \sinh(s)  \;, \label{boosts}
\end{equation}
a spinor field $\psi$ undergoes a transformation given by 
\begin{equation} 
{\cal U}(\Lambda_s) \psi(\theta) {\cal U}^{-1}(\Lambda_s) = e^{-\frac{s}{2} \gamma^0 \gamma^1} \psi(\theta-s)  \;, \label{btr}
\end{equation}
it follows that the combination ${\hat f}(\theta)$, eq.\eqref{comb}, transforms precisely as a scalar, due to the presence of $e^{-\frac{\theta}{2}}$ and $e^{\frac{\theta}{2}}$, namely 
\begin{eqnarray} 
{\cal U}(\Lambda_s) {\hat f}(\theta) {\cal U}^{-1}(\Lambda_s)& =& e^{-\frac{\theta}{2}} {\cal U}(\Lambda_s) { f_1}(\theta) {\cal U}^{-1}(\Lambda_s) + i e^{\frac{\theta}{2}} {\cal U}(\Lambda_s) { f_2}(\theta) {\cal U}^{-1}(\Lambda_s) \nonumber \\
&=& e^{-\frac{\theta-s}{2}} f_1(\theta-s) + i e^{\frac{\theta-s}{2}}f_2(\theta-s) = {\hat f}(\theta-s) 
\end{eqnarray}
in agreement with the scalar nature of ${\bar f}\psi$ in expression \eqref{sm}. \\\\The smeared operators $(b_f,b^\dagger_f)$ obey the following anti-commutation relations 
\begin{equation}
\{ b_f, b^\dagger_g \} = \frac{m}{2} \int_{-\infty}^{\infty} \frac{d\theta}{2 \pi} {\hat f}^*(\theta) {\hat g}(\theta) \;, \qquad \{ b_f, b_g \} =\{ b^\dagger_f, b^\dagger_g \} = 0 \;, \label{smbb}
\end{equation}
where $({\hat f}(\theta), {\hat g}(\theta))$ stand for the scalar combinations of eq.\eqref{comb}. Let us now look at the Wightman two point function $\langle 0| {\psi_f} {\psi_g} |0\rangle$ which, by construction, defines the inner product in the 1-particle Hilbert space of the theory \cite{Haag:1992hx}, {\it i.e} 
\begin{equation} 
\langle 0| {\psi_f} {\psi_g} |0\rangle  = \langle f| g \rangle \;, \label{innp}
\end{equation} 
where 
\begin{equation} 
b_f |0\rangle = 0 \;, \forall f \;. \label{vacuum}
\end{equation}
Expression \eqref{innp} gives\footnote{The presence of the factor $m/2$ is just a matter of convention. It might be reabsorbed by redefining the normalization of the spinors $(u(\theta),v(\theta))$ in eqs.\eqref{uth},\eqref{vth}.}
\begin{equation} 
\langle f |g \rangle = \frac{m}{2} \int \frac{ d \theta}{2 \pi} \left( e^{-\frac{\theta}{2}} f^*_1 - i e^{\frac{\theta}{2}} f^*_2 \right) \left( e^{-\frac{\theta}{2}} g_1 + i e^{\frac{\theta}{2}} g_2 \right) =  \frac{m}{2} \int \frac{ d \theta}{2 \pi} {\hat f}^*(\theta) \;{\hat g}(\theta)  \;. \label{innfin}
\end{equation}
The inner product $\langle f|g\rangle$ depends only on the scalar combinations $({\hat f},{\hat g})$. It turns out thus that the 1-particle Hilbert space of the theory is 
\begin{equation} 
{\cal H}_{1-part} = L^2({\mathbb{R}},d\theta)  \;, \label{L2}
\end{equation}
as a complex Hilbert space equipped with the Majorana real structure. The wedge-localized spaces introduced below are real standard subspaces of this complex 1-particle Hilbert space. This reflects the physical fact that the massive Majorana field in $1+1 $ describes only one type of particle, as exhibited by the presence of a single operator $b(\theta)$ in the plane wave expansion \eqref{wexp}. This feature stems from the Majorana condition and from the fact that the two components $(h, \varphi)$ get mixed by the non-vanishing mass term. Let us conclude this section by showing the positivity of the norm: 
\begin{equation} 
||f||^2 = \frac{m}{2} \int \frac{ d \theta}{2 \pi} \left( e^{-\frac{\theta}{2}} f^*_1 - i e^{\frac{\theta}{2}} f^*_2 \right) \left( e^{-\frac{\theta}{2}} f_1 + i e^{\frac{\theta}{2}} f_2 \right) =  \frac{m}{2} \int \frac{ d \theta}{2 \pi} {\hat f}^*(\theta) \;{\hat f}(\theta)  \ge 0\;. \label{norm}
\end{equation}

\section{Tomita-Takesaki modular theory. The Bell-CHSH inequality in the vacuum state }\label{sect4}

We have now all ingredients to formulate the Bell-CHSH inequality for wedge regions $(W_R,W_L)$, defined as 

\begin{equation} 
W_R =\{(x,t), \; x> |t|\} \;, \qquad  W_L =\{(x,t), \; -x< |t|\} \;. \label{wrwl}
\end{equation}

These regions are the causal complements of each other. An important feature of the wedges $(W_R,W_L)$ is that they are left invariant by the boost transformations, a basic property at the heart of the Bisognano-Wichmann results \cite{Bisognano:1975ih}. These authors have been able to give a full characterization of both modular operator $\delta$ and modular conjugation $j$ in wedge regions. The operator $\delta$ is self-adjoint and positive definite, being given by 
\begin{equation} 
\delta = e^{-2 \pi K} \;, \qquad K = -i \frac{d}{d\theta} \label{modd}
\end{equation}
where $K$ is the self-adjoint boost generator. On the other hand, the modular conjugation $j$ is anti-unitary: 
\begin{equation} 
j = R_3(\pi) (CPT) \;, \label{modj}
\end{equation}
where $(CPT)$ stands for the CPT operator and $R_3(\pi)$ is a rotation of $\pi$ around the x-axis. From $K=-i \frac{d}{d \theta}$ one sees that the spectrum of $K$ is the whole real axis $\mathbb{R}$, {\it i.e.}
\begin{equation} 
K \psi_\omega = \omega \psi_\omega \;, \qquad \omega \in \mathbb{R} \;, \label{omeig}
\end{equation}
with generalized eigenstates $\psi_\omega$ given by plane waves in rapidity space 
\begin{equation} 
\psi_\omega = \frac{1}{\sqrt{2\pi}} e^{i \omega \theta} \;, \qquad \langle \psi_\omega| \psi_{\omega'} \rangle = \frac{1}{2 \pi} \int_{-\infty}^\infty d \theta e^{-i \theta(\omega - \omega')} = \delta(\omega-\omega') \;. \label{eig}
\end{equation}
It follows thus that the spectrum of $\delta$ is continuous, its eigenvalues $\lambda^2(\omega)$ being 
\begin{equation} 
\lambda^2(\omega) = e^{- 2 \pi \omega} \;, \qquad \lambda^2 \in \mathbb{R}_+ \;. \label{leig}
\end{equation}
Following \cite{Guido:2008jk}, out of the operators $\delta$ and $j$, one introduces the unbounded anti-linear Tomita-Takesaki operator $s$: 
\begin{equation} 
s = j \delta^{1/2} \;. \label{s-op}
\end{equation}
The operators $(s,j,\delta)$ enjoy the following properties \cite{Guido:2008jk}:
\begin{equation} 
s^2=1 \; \qquad j^2=1 \;, \qquad j \delta^{1/2} j = \delta^{-1/2} \;, \qquad s^\dagger s = \delta \;. \label{prop}
\end{equation}
When acting on elements $\{ \psi(\theta) \}$ of the 1-particle Hilbert space $L^2({\mathbb R},d\theta)$, the operator $\delta^{1/2}$ and $j$ give, respectively, 
\begin{equation} 
\delta^{1/2} \psi(\theta) = \psi(\theta+i \pi) \;, \qquad j\psi(\theta) = \psi(\theta)^*  \;, \label{djact}
\end{equation}
so that 
\begin{equation} 
s \psi(\theta) = \psi(\theta+i\pi)^* \;. \label{sact}
\end{equation}
One sees that, in order to have well defined modular operators, the requirement that $\{ \psi(\theta) \}$ exhibit a bounded analytic extension in the strip $(\theta+iy), \; 0\le y\le \pi$,  is needed. \\\\The Tomita-Takesaki operator $s$ plays a central role in the so-called modular localization  \cite{Brunetti:2002nt,Borchers:2000pv,Schroer:2014kya}, a very beautiful and powerful tool for analyzing the Bell-CHSH inequality. To that end, one introduces the real subspace  \cite{Guido:2008jk} $K(W_R)$ defined as the closure of 
\begin{equation} 
K(W_R) \equiv  \{ \psi_\xi(\theta) = \int d^2x e^{-i p^\mu(\theta) x_\mu} \xi(t,x) \;, \; supp(\xi) \in W_R \}
\end{equation}
where $\xi(t,x)$ is a smooth function supported in the wedge region $W_R$. By construction, $\psi_\xi(\theta)$ exhibits an analytic continuation  in the strip $\theta +iy, \; 0\le y\le \pi$, see \cite{Caribe:2026mam}. The relevance of the real subspace $K(W_R)$ relies on the fact that it is a standard subspace for the 1-particle Hilbert space, meaning that \cite{Guido:2008jk}
\begin{itemize} 
\item $K(W_R) \cap i K(W_R) = \{ 0 \}$ 

\item $K(W_R) + i K(W_R)$ {\it  \;\;\;dense in the  1-particle Hilbert space }
\end{itemize}
These important properties state that any vector of the 1-particle Hilbert space can be arbitrarily well approximated by elements belonging to $K(W_R)$ and $iK(W_R)$ and which, in turn, originate from smooth functions localized in the wedge $W_R$. \\\\For further use, it is helpful to introduce the symplectic complement $K'(W_R)$ defined as 
\begin{equation} 
K'(W_R) = \{ \eta \in L^2({\mathbb R},d\theta), \; \Im \langle \eta|\zeta \rangle = 0 \;\;\; \forall \zeta \in K(W_R) \}
\end{equation}
Elements of $K'(W_R)$ are space-like with respect to elements of $K(W_R)$  \cite{Guido:2008jk,Caribe:2026mam}. Moreover, the Tomita-Takesaki theory states that  \cite{Guido:2008jk}
\begin{equation} 
K'(W_R) = j K(W_R)  \;. \label{ttth}
\end{equation}
Also, from the Haag duality \cite{Bisognano:1975ih,Guido:2008jk} one has\footnote{Given an open bounded region ${\cal O }$ of the Minkowski spacetime, its causal complement $ {\cal O}'$ is given by 
\begin{equation} 
{\cal O}'= \{ (t,x) \in {\mathbb{R}}^2, \; (x-y)^2<0 \;\;\;\forall y \in {\cal O} \} \;. \label{causc}
\end{equation}
For the wedge regions $(W_R,W_L)$, it holds that $W'_R=W_L$. }
\begin{equation} 
K'(W_R) = K(W'_R)=K(W_L) \;. \label{dual}
\end{equation}
From  \cite{Brunetti:2002nt,Borchers:2000pv,Schroer:2014kya}, one learns that the standard subspace $K(W_R)$ can be defined entirely in terms of the Tomita-Takesaki operator $s$, namely 
\begin{equation} 
K(W_R) =\{ \psi(\theta) \in L^2({\mathbb R},d\theta), \; s\psi(\theta) = \psi(\theta) \} \;. \label{swr}
\end{equation}
Similarly 
\begin{equation} 
K'(W_R) =\{ \eta(\theta) \in L^2({\mathbb R},d\theta), \; s^\dagger\eta(\theta) = \eta(\theta) \} \;. \label{sdwr}
\end{equation}
Equation \eqref{swr},\eqref{sdwr} express the content of the modular localization. One says that an element of the Hilbert space is $W_R$-localized if 
\begin{equation} 
s \psi(\theta) = \psi(\theta) \qquad W_R-localized  \;. \label{wrl}
\end{equation}
In the same way, a vector $\eta$ is $W_L$-localized if 
\begin{equation} 
s^\dagger \eta(\theta) = \eta(\theta) \qquad W_L-localized \;. \label{wll}
\end{equation}
In particular, from eq.\eqref{wrl}, the condition for $W_R$-localization becomes 
\begin{equation} 
s \psi(\theta) = \psi(\theta) \Rightarrow \psi(\theta) = \psi(\theta+ i \pi)^* \;. \label{kms}
\end{equation}
We are now ready to formulate the Bell-CHSH inequality in the vacuum state. The first task is that of introducing a genuine dichotomic Hermitian field operator ${\cal A}(f)$. In the case of Fermi fields, this goal is achieved by employing the spinor field $\psi(f)$ itself \cite{Summers:1987fn,Summers:1987squ}. In fact, from eqs.\eqref{smbb}, it follows that the operator 
\begin{equation} 
{\cal A}(f) = \frac{\psi(f)}{||f||} \;, \label{Af}
\end{equation}
is Hermitian and dichotomic 
\begin{equation} 
{\cal A}(f)= {\cal A}(f)^\dagger \;, \qquad {\cal A}(f)^2=1 \;. \label{Afd}
\end{equation}

There is one fermionic point which has to be kept explicit. If $f$ and $g$ are localized in opposite wedges, the odd Majorana operators ${\cal A}(f)$ and ${\cal A}(g)$ graded-commute rather than commute. To obtain ordinary commuting Bell observables, we use the usual fermion-parity twist. Let $N$ denote the fermion number operator and
\begin{equation}
\Gamma=(-1)^N \;, \qquad \Gamma |0\rangle = |0\rangle \;, \qquad \Gamma {\cal A}(f)\Gamma=-{\cal A}(f) \;,
\end{equation}
and define the left-wedge, Klein-transformed observables
\begin{equation}
{\cal B}(g)= i\Gamma {\cal A}(g) \;, \qquad {\cal B}(g')= i\Gamma {\cal A}(g') \;. \label{Btwist}
\end{equation}
They are Hermitian and dichotomic,
\begin{equation}
{\cal B}(g)^\dagger={\cal B}(g) \;, \qquad {\cal B}(g)^2=1 \;,
\end{equation}
and they commute with the right-wedge observables ${\cal A}(f)$ and ${\cal A}(f')$. Thus the following is an ordinary Bell-CHSH operator built from commuting operators attached to the two wedges. In vacuum expectation values this convention gives
\begin{equation}
\langle 0|{\cal A}(f){\cal B}(g)|0\rangle
= -i \langle 0|{\cal A}(f){\cal A}(g)|0\rangle \;,
\label{twistcorr}
\end{equation}
which is the origin of the factor $-i$ below.

As such, for the Bell-CHSH in the vacuum state $|0\rangle$ one has 
\begin{equation} 
\langle 0| {\cal C} |0\rangle = \langle 0| \left( {\cal A}(f) + {\cal A}(f') \right) {\cal B}(g) +  \left( {\cal A}(f) - {\cal A}(f') \right) {\cal B}(g') |0\rangle  \;, \label{bbll}
\end{equation}
where $(f,f')$ and $(g,g')$ are, respectively, $W_R$ and $W_L$ localized, {\it i.e.} 
\begin{equation} 
sf = f \;, \qquad sf'=f'\;, \qquad s^\dagger g= g \;, \qquad s^\dagger g'= g' \;. \label{ffgg}
\end{equation}
Making use of the inner product, eqs.\eqref{innp},\eqref{innfin}, the Bell-CHSH correlator becomes 
\begin{equation} 
\langle 0| {\cal C} |0\rangle = -i \left(      \frac{\langle f|g\rangle}{||f|| ||g||}  +   \frac{\langle f'|g\rangle}{||f'|| ||g||}    +   \frac{\langle f|g'\rangle}{||f|| ||g'||}  -   \frac{\langle f'|g'\rangle}{||f'|| ||g'||}  \right) \;, \label{bf1}
\end{equation}
where the factor $-i$ is precisely the twist factor displayed in eq.\eqref{twistcorr}. The Bell-CHSH inequality is said to be violated in the vacuum state $|0\rangle$ whenever 
\begin{equation} 
2 < | \langle 0| {\cal C} |0\rangle | \le 2\sqrt{2} \;. \label{vv}
\end{equation}

\section{Approach to the Tsirelson bound }\label{sect5}

This section is devoted to present our analytic setup for the Bell-CHSH correlator, eq.\eqref{bf1}. We begin with the modular localization of $(f,f')$ and $(g,g')$. To that end we pick up a function $\phi(\theta)$ exhibiting analytic extension in the complex upper half-plane $\theta+iy,\; y \ge 0$\footnote{Of course, this region includes the strip $\theta +iy,\; 0\le y\le \pi$.}, so as to ensure that the Tomita-Takesaki operator $s$ is well defined. This requirement can be achieved by making use of the half-sided Fourier transformation, namely 
\begin{equation} 
\phi(\theta) = \int_0^\infty d\omega\; h(\omega) e^{i \omega \theta}  \;, \label{halfs}
\end{equation}
where $h(\omega)$ is a smooth real function satisfying
\begin{equation}
\int_0^\infty d\omega\; h^2(\omega)e^{2\pi\omega}<\infty \;.
\label{hadm}
\end{equation}
This condition ensures that $h(\omega)$ decays sufficiently fast to justify the boundary value of $\phi(\theta)$ at $\theta+i\pi$.

It is apparent that expression \eqref{halfs} meets the desired analytic requirement. It should be observed that eq.\eqref{halfs} can be seen as the expansion of $\phi(\theta)$ in the rapidity plane waves $\{ e^{i \omega \theta} \}$ which are precisely the generalized eigenstates of the operator $\delta$, eqs.\eqref{omeig}-\eqref{leig}. As such, eq.\eqref{halfs}  has the meaning of an expansion along the spectrum of the modular operator $\delta$, which is nothing but  the half-line ${\mathbb{R}}_+$.
Following \cite{Summers:1987fn,Summers:1987squ},  the  modular localization of the vectors $(f,f')$ and $(g,g')$ is achieved in two steps. First, one sets 
\begin{eqnarray} 
f &=& (1+s) \phi \;, \qquad f'= (1+s) i \phi \; \nonumber \\
{\tilde g} &=& i (1+ s^\dagger) \phi \;, \qquad {\tilde g'} = - i (1+ s^\dagger) i \phi  \;. \label{swcc}
\end{eqnarray}
Since $s^2=1$, it turns out that, as required, $(f,f')$ are $W_R$-localized while $({\tilde g},{\tilde g'})$ are $W_L$-localized, {\it i.e} 
\begin{equation} 
sf=f \;, \qquad sf'= f'\;, \qquad s^\dagger {\tilde g} = {\tilde g} \;, \qquad s^\dagger {\tilde g'} = {\tilde g'} \;. \label{locffgt}
\end{equation}
A simple calculation shows that 
\begin{eqnarray} 
f & =& \int_0^\infty d\omega \; h(\omega) \left( e^{i \omega \theta} + e^{-i \omega \theta} e^{-\pi \omega} \right) \;, \nonumber \\
f' & =& i \int_0^\infty d\omega \; h(\omega) \left( e^{i \omega \theta} - e^{-i \omega \theta} e^{-\pi \omega} \right) \;, \nonumber \\
{\tilde g} & =& i \int_0^\infty d\omega \; h(\omega) \left( e^{i \omega \theta} + e^{-i \omega \theta} e^{\pi \omega} \right) \;, \nonumber \\
{\tilde g'} & =&  \int_0^\infty d\omega \; h(\omega) \left( e^{i \omega \theta} - e^{-i \omega \theta} e^{\pi \omega} \right) \;. \label{fourex}
\end{eqnarray}
These expressions give rise to the following inner products: 
\begin{eqnarray} 
\langle f | {\tilde g}\rangle &=& 2 (2\pi) i \int_0^\infty d\omega \; h^2(\omega) = - \langle f' |{\tilde g'}\rangle \;, \nonumber \\
\langle f | {\tilde g'} \rangle & = & \langle f'| {\tilde g} \rangle = 0 \;, \nonumber \\
\langle f | f'\rangle &=& (2\pi)i \int_0^\infty d\omega\; h^2(\omega) \left( 1 - e^{-2\pi \omega} \right) = - \langle {\tilde g} | {\tilde g'} \rangle \;. \label{innrs}
\end{eqnarray}
The final form of $(g,g')$ is then obtained by defining \cite{Summers:1987fn,Summers:1987squ} 
\begin{equation} 
g = \frac{{\tilde g} - {\tilde g'}}{\sqrt{2}} \;, \qquad g' = \frac{{\tilde g} + {\tilde g'}}{\sqrt{2}} \;. \label{fggp}
\end{equation}
Therefore, noticing that 
\begin{eqnarray} 
||f||^2 &=& ||f'||^2 = (2\pi) \int_0^\infty d\omega \; h^2(\omega) \left(1 + e^{-2\pi \omega} \right) \;, \nonumber \\
||g||^2 &=& ||g'||^2 = (2\pi) \int_0^\infty d\omega \; h^2(\omega) \left(1 + e^{2\pi \omega} \right) \;, \label{nff}
\end{eqnarray}
for the Bell-CHSH correlator we get the compact expression 
\begin{equation} 
\langle 0| {\cal C} |0\rangle = 2\sqrt{2} \frac{2 \int_0^\infty d\omega \; h^2(\omega)  }{\sqrt{\int_0^\infty d\omega \; h^2(\omega)\left(1 + e^{-2\pi \omega} \right) }{\sqrt{\int_0^\infty d\omega \; h^2(\omega)\left(1 + e^{2\pi \omega} \right) }} } \;. \label{bellf1}
\end{equation} 

For admissible nonzero regular weights, eq.\eqref{bellf1} gives a value below the Tsirelson bound. To see this, set $\lambda^2(\omega)=e^{-2\pi\omega}$. Then
\begin{equation}
\left(\int_0^\infty d\omega\; h^2(1+\lambda^2)\right)
\left(\int_0^\infty d\omega\; h^2(1+\lambda^{-2})\right)
\ge
\left(\int_0^\infty d\omega\; h^2\sqrt{(1+\lambda^2)(1+\lambda^{-2})}\right)^2
\ge
4\left(\int_0^\infty d\omega\; h^2\right)^2 \;.
\label{csbound}
\end{equation}
Equality requires the spectral weight to be supported at $\omega=0$. Thus $2\sqrt{2}$ is the supremum reached by concentrating $h^2(\omega)$ near $\omega=0$, while every regular admissible weight gives a strictly smaller value.

Expression \eqref{bellf1} exhibits several features, worth to be underlined, {\it i.e.} 
\begin{itemize} 
\item with the choice of $\phi(\theta)$ as given in eq.\eqref{halfs}, the inner products entering the Bell-CHSH inequality read
\begin{equation} 
 \frac{\langle f|g\rangle}{||f|| ||g||}= \frac{\langle f'|g\rangle}{||f'|| ||g||}= \frac{\langle f|g'\rangle}{||f|| ||g'||}=- \frac{\langle f'|g'\rangle}{||f'|| ||g'||}= \frac{\sqrt{2} \int_0^\infty d\omega \; h^2(\omega)  }{\sqrt{\int_0^\infty d\omega \; h^2(\omega)\left(1 + e^{-2\pi \omega} \right) } {\sqrt{\int_0^\infty d\omega \; h^2(\omega)\left(1 + e^{2\pi \omega} \right) }} }  \;. \label{iinnpp}
\end{equation} 
In particular, it should be noted that $\frac{\langle f'|g'\rangle}{||f'|| ||g'||}$ has opposite sign with respect to the other inner products, the sign pattern needed to approach the Tsirelson value. This property, already present in the original work by Summers-Werner \cite{Summers:1987fn,Summers:1987squ}, has been proven also in the recent paper by \cite{Dudal:2026eil}. 

\item Expression \eqref{bellf1} also enables one to keep track of the eigenvalues of the modular operator $\delta$ in a quite clear way. For that, it suffices to recast eq.\eqref{bellf1} in the form 
\begin{equation} 
\langle 0| {\cal C} |0\rangle = 2\sqrt{2} \frac{2 \int_0^\infty d\omega \; h^2(\omega)  }{\sqrt{\int_0^\infty d\omega \; h^2(\omega)\left(1 + \lambda^2(\omega) \right) }{\sqrt{\int_0^\infty d\omega \; h^2(\omega)\left(1 + \lambda^{-2}(\omega) \right) }} } \;. \label{bellf2}
\end{equation}
where $\{ \lambda^2(\omega) = e^{-2\pi \omega}\}$ denote the eigenvalues of $\delta$. The near-Tsirelson regime is obtained when the spectral weight is concentrated near $\lambda=1$, or equivalently near $\omega=0$. This point belongs to the continuous modular spectrum rather than to an isolated normalizable eigenvector. The connection with the type $III_1$ nature of local relativistic algebras should therefore be read spectrally: the relevant modular spectrum accumulates at $\lambda=1$, and the Bell-CHSH supremum is approached by weights concentrated there. From that perspective, we expect that any choice of the function $h(\omega)$ which implements an effective projection to the value $\omega\approx 0$, {\it i.e.} $\lambda(\omega)\approx 1$, will lead to arbitrarily close approach to the Tsirelson bound. That this is indeed the case, will be discussed in the next subsection. 

\end{itemize}

\subsection{Analytic solutions}\label{an}

\begin{figure}[t!]
\begin{minipage}[b]{0.4\linewidth}
\includegraphics[width=\textwidth]{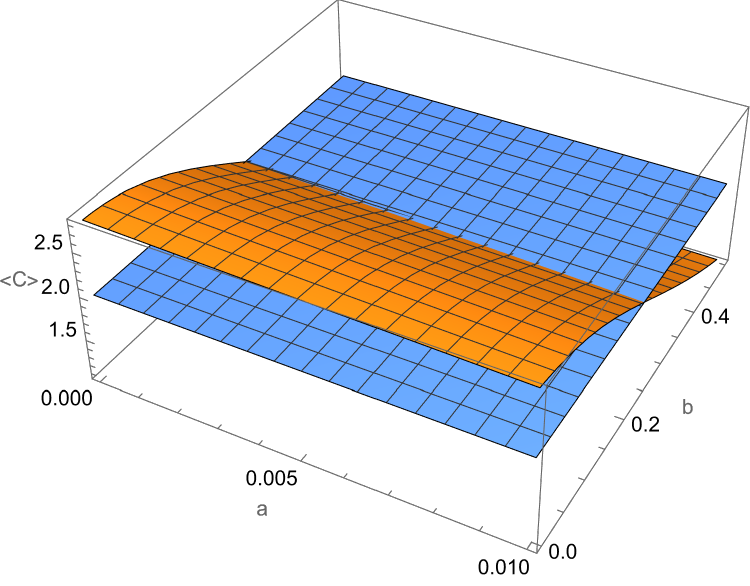}
\end{minipage} \hfill
 \caption{Behavior of the Bell-CHSH correlator $\langle 0 |{\cal C} |0\rangle$ as a function of the parameters $(a,b)$. The blue surface denotes the classical bound: 2. The orange surface above the blue one corresponds to the values of $(a,b)$ for which violations of the Bell-CHSH inequality take place. Near-Tsirelson values occur for small values of $b$ and $a \approx 0$, where the spectral weight is concentrated near $\omega=0$.}
\label{Fig1}
\end{figure}

We present here a few analytic families whose Bell-CHSH values approach the Tsirelson bound. The first and, perhaps, simplest choice for $h(\omega)$ is a two parameter Gaussian: 
\begin{equation} 
h(\omega) = {\cal N} e^{-\frac{(\omega-b)^2}{2a} }\;, \label{gauss}
\end{equation}
with ${\cal N}$ a normalization constant and $(a,b)$ two free positive parameters. One notices that, for very small values of $a$, expression \eqref{gauss} is essentially different from zero only when $\omega \approx b$. As such, according to the previous considerations,near-maximal violation is expected for small values of $b$ and $a\approx0$. From 
\begin{eqnarray} 
\int_0^\infty d\omega \; e^{-\frac{(\omega-b)^2}{a} }\left( 1 + e^{2\pi \omega} \right) &=& \frac{\sqrt{\pi a}}{2} \left(  1 + erf\left( \frac{b}{\sqrt{a}} \right) + e^{2\pi b} e^{\pi^2 a} \left(1 + erf\left( \frac{b+ \pi a}{\sqrt{a}} \right)  \right)    \right) \;, \nonumber \\
\int_0^\infty d\omega \; e^{-\frac{(\omega-b)^2}{a} }\left( 1 + e^{-2\pi \omega} \right) &=& \frac{\sqrt{\pi a}}{2} \left(  1 + erf\left( \frac{b}{\sqrt{a}} \right) + e^{-2\pi b} e^{\pi^2 a} \left(1 + erf\left( \frac{b- \pi a}{\sqrt{a}} \right)  \right)    \right) \;, \nonumber \\
\int_0^\infty d\omega \; e^{-\frac{(\omega-b)^2}{a} } & =& \frac{\sqrt{\pi a}}{2} \left( 1 + erf\left( \frac{b}{\sqrt{a}} \right) \right)\;, \label{int}
\end{eqnarray} 
with $erf(x)$ denoting the error function 
\begin{equation}
erf(x) = \frac{2}{\sqrt{\pi}}\int_0^x dt \; e^{-t^2} \;, \label{error}
\end{equation} 
for the Bell-CHSH inequality one gets the closed expression 
\begin{equation} 
\langle 0| {\cal C} |0\rangle = \frac{4 \sqrt{2} \left( 1 + erf\left( \frac{b}{\sqrt{a}} \right) \right)} {\Delta_+^{1/2} \Delta_-^{1/2}} \;, \label{blln1}
\end{equation} 
with 
\begin{equation} 
\Delta_{\pm}= 1 + erf\left( \frac{b}{\sqrt{a}} \right) + e^{\pm 2\pi b} e^{\pi^2 a} \left(1 + erf\left( \frac{b\pm \pi a}{\sqrt{a}} \right)  \right) \;. \label{Dpm}
\end{equation} 
The behavior of $\langle 0| {\cal C} |0\rangle$ as a function of the parameters $(a,b)$ is shown in Fig.\eqref{Fig1}. The blue surface stands for the classical value: 2. The orange surface above the blue one gives the region in the parameter space $(a,b)$ for which the violation of the Bell-CHSH inequality takes place. As expected, the size of the violation increases when the parameter $a$ approaches zero and $b$ becomes small. For these values, the Gaussian \eqref{gauss} implements a sharp concentration near $\omega \approx 0$, {\it i.e.} near $\lambda\approx 1$. In fact, it turns out that 
\begin{equation} 
\langle 0 |{\cal C} |0\rangle = 2.82842 \;, \qquad a \approx 10^{-6} \;, \qquad b \approx 0.0001 \;, \label{tssi1}
\end{equation}
showing a near-Tsirelson value. The exact value $2\sqrt{2}$ is recovered in the limiting sense described after eq.\eqref{bellf1}.

\begin{figure}[t!]
\begin{minipage}[b]{0.4\linewidth}
\includegraphics[width=\textwidth]{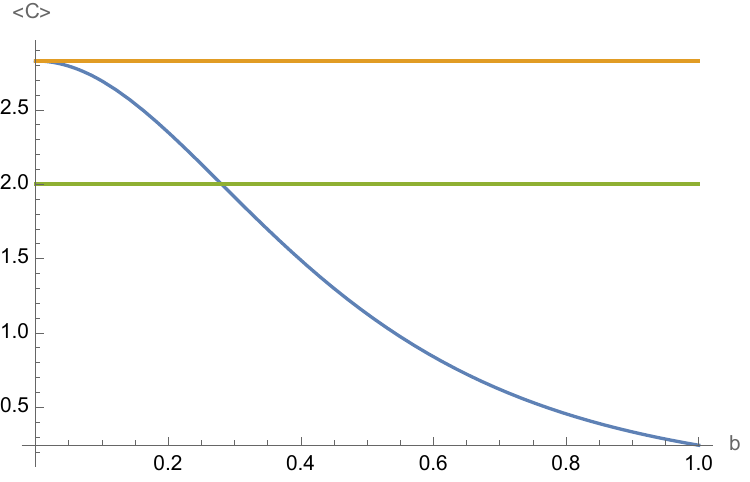}
\end{minipage} \hfill
 \caption{Behavior of the Bell-CHSH correlator $\langle 0 |{\cal C} |0\rangle$ as a function of the parameter $b$.  Large violations take place for small values of $b$, with the Tsirelson value approached as $b\to0$. }
\label{Fig2}
\end{figure}

\noindent A second, one parameter example, is provided by the Lorentzian type expression 
\begin{equation} 
h^2(\omega) = \frac{\varepsilon e^{-2\pi\omega}}{(\omega -b)^2 + \varepsilon} \;, \label{lz}
\end{equation}
with $\varepsilon$ infinitesimal. In this case, for the Bell-CHSH inequality one gets 
\begin{equation} 
\langle 0 |{\cal C} |0\rangle = 4 \sqrt{2} \frac{1}{\sqrt{\left(1+ e^{-2\pi b}\right)\left(1+e^{2 \pi b}\right)}} = \frac{2\sqrt{2}}{\cosh(\pi b)} \; \label{Lzv}
\end{equation}
whose behavior is reported in Fig.\eqref{Fig2}. 

\vspace{3mm}
\noindent Again, the approach to the Tsirelson bound occurs for $b \approx 0$, corresponding to $\lambda^2(b) = e^{-2\pi b} \approx 1$. \\\\It becomes apparent now that many analytic families can be found in a rather simple and quick way. Though, as far as near-saturation of the Tsirelson bound is concerned, only the region for which $\lambda \approx 1$ matters, as expressed by eq.\eqref{bellf2}.

\section{Conclusion}\label{sect6}

In this work, the violation of the Bell-CHSH inequality in the vacuum state has been examined for the massive Majorana field in $1+1$ dimensions. The anti-commutation relations give normalized Hermitian dichotomic odd field operators, while the fermion-parity twist in eq.\eqref{Btwist} turns the left-wedge operators into commuting Bell partners for the right-wedge operators.

Using the Bisognano-Wichmann theorem and modular wedge localization, we derived a rapidity-space formula for the vacuum Bell-CHSH correlator. The main result is eq.\eqref{bellf1}: the correlator depends only on one admissible spectral weight $h^2(\omega)$ for the modular operator. For any regular admissible $h$, the value is below $2\sqrt{2}$ with the Tsirelson value being the supremum reached by spectral concentration near $\omega=0$, or equivalently near $\lambda^2(\omega)=1$.

The analytic Gaussian and Lorentzian-type examples, eqs.\eqref{blln1} and \eqref{Lzv}, show that the Tsirelson bound, {\it i.e.} $2\sqrt{2}$, is approached whenever one gets close to the eigenvalue $\lambda\approx 1$ of the Tomita-Takesaki modular operator $\delta$. This is not a mere coincidence. In fact, as pointed out in \cite{Summers:1987fn,Summers:1987squ}, this value corresponds to the fixed point of the modular flow $\delta^{it} =e^{it \log(\delta)}$, being connected to the important fact that the von Neumann local algebraic structure of a relativistic Quantum Field Theory is of the type $III_{\lambda=1}$. 

The present setup gives a concrete Majorana-field realization of the Summers-Werner mechanism and may be useful for future investigations of the Bell-CHSH inequality in $1+2$ and $1+3$ dimensions for both massive and massless fields. The possibility of addressing the Mermin inequalities can be also envisaged. 

Finally, we point  out that Majorana excitations are a hot topic in condensed matter systems, from both theoretical as well as experimental sides. One might quote: the fractional quantum Hall edge states, the Luttinger liquids in quantum wires, the anyonic interferometry, the cold atoms in one dimension, etc, see, for instance, ref.\cite{nat,Rachel:2024gjf,Caribe:2026xoa} for a recent account. As such, the present contributions could open interesting perspectives to devise Bell tests in such a systems.

\section*{Acknowledgments}
The authors would like to thank the Brazilian agencies CNPq, CAPES and FAPERJ for financial support.  S. P.~Sorella, I.~Roditi, and M. S.~Guimaraes are CNPq researchers under contracts 302991/2024-7, 311876/2021-8, and 309793/2023-8, respectively. Prof. Ricardo Correa da Silva is gratefully acknowledged for fruitful discussion. 


\end{document}